\documentclass[twocolumn,floatfix,pre,aps,showpacs]{revtex4}
\usepackage{graphicx}
\usepackage{amsmath,mathrsfs,epsfig}

\begin{document}

\title{Critical Lattice Size Limit for Synchronized Chaotic State in 1-D and 2-D
Diffusively Coupled Map Lattices}

\author{P. Palaniyandi}
\email{palani@math.iisc.ernet.in}
\affiliation{Department of Mathematics, Indian Institute of Science,
Bangalore 560 012, India}
\author{Govindan Rangarajan}
\email{rangaraj@math.iisc.ernet.in}
\affiliation{Department of Mathematics, Indian Institute of Science,
Bangalore 560 012, India}

\date{\today}

\begin{abstract}
We consider diffusively coupled map lattices with $P$ neighbors (where $P$ is
arbitrary) and study the stability of synchronized state.  We show that
there exists a critical lattice size beyond which the synchronized state is
unstable. This generalizes earlier results for nearest neighbor coupling.
We confirm the analytical results by performing numerical simulations on
coupled map lattices with logistic map at each node. The above analysis is also
extended to $2$-dimensional $P$-neighbor diffusively coupled map lattices.
\end{abstract}


\maketitle

\section{Introduction}
\label{intro}

In recent years, synchronization of coupled dynamical
systems~\cite{pecora:90:01, pecora:91:01, pikovsky:book:01:01} has become an
important area of research for their applications in a variety of fields
including secure communications, cryptography, optics, neural networks, pattern
formation, geophysics and population dynamics~\cite{cuomo:93:01,
palaniyandi:01:01, rangarajan:03:01, amritkar:06:01}. In particular,
the stability of synchronized state in coupled map lattices (CML) with various
coupling schemes have been studied extensively\cite{anteneodo:03:01,
ginelli:03:01, cencini:05:01, heagy:94:01, brown:97:01,
glendinning:99:01, rulkov:92:01, kaneko:84:01,
belykh:00:01, belykh:01:01, zhan:00:01, rangarajan:02:01,
chen:06:01,chen:03:01, pecora:98:01, palaniyandi:05:03}. To be specific the CML
with diffusively coupling has attracted considerable attention in recent
studies. In such systems, the synchronized state is not stable when the number
of nodes exceeds a certain critical limit and each node is coupled only with
its nearest neighbor~\cite{rangarajan:02:01,chen:03:01}. In this paper, using the formalism put forth
in Refs. \cite{rangarajan:02:01,chen:03:01} we derive an
exact analytic expression for this limit for more general case of $P$-neighbor
coupling.   Further, the results are verified through numerical simulations in
coupled logistic map lattices.  All the analysis are carried out in both 1D and
2D CMLs. Studies similar to our present work, but for coupled oscillators, are
reported in Ref.~\cite{barahona:02:01,belykh:04:01}.

\section{Critical Size Limit in 1D case}
\label{stability_1d}
Consider 1-dimensional coupled map lattices with $P$-neighbor
diffusive coupling represented by
\begin{align}
{\bf x}_j(n+1) =  f\big({\bf x}_j(n)\big) +  \frac{1}{2P} \sum^P_{p=1} a_p \left[ f\big({\bf x}_{j-p}(n)\big) + \right . \nonumber \\
\left. f\big({\bf x}_{j+p}(n)\big) - 2f\big({\bf x}_j(n)\big) \right],
\label{cml_1d}
\end{align}
where ${\bf x}_j$ is a $M$-dimensional state vector, $j$  represents the
lattice site, $L$ is the lattice size, $a_p$ is the coupling strength between
$j$th map and its $p$th neighbor, and  the evolution of the map at $j$th site
is described by $f\big({\bf x}_j(n)\big)$. Also periodic boundary condition is
imposed and the synchronized state (synchronization manifold) is defined by
${\bf x}_1(n)={\bf x}_2(n)=\dots={\bf x}_L(n)={\bf x}(n)$. Since $a_p$ is a very
general coupling coefficient, the long range model proposed by Antenedo \cite{anteneodo:03:01}
can be incorporated into the above equation.

Linearizing (\ref{cml_1d}) around ${\bf x}$ and performing the discrete spatial
Fourier transform ${\bf \eta}_l(n)= \frac{1}{L}\sum^L_{j=1}   \text{exp}(-i2\pi
j l/L){\bf z}_j(n)$, the resulting form after simplification (see
Refs.~\cite{rangarajan:03:01, chen:06:01} for details) is
\begin{align}
\label{trans_lya_1d}
\mu_i(l)=h_i+ln\left|1-\frac{2}{P}\sum^P_{p=1}a_p\text{sin}^2(\pi pl/L)\right|, \\
i=1,2,\dots,M; \ \ l=0,1, \ldots , L-1.  \nonumber
\end{align}
Here $\mu_i(l)$'s are the Lyapunov exponents corresponding to $l$th mode and
$h_i$'s are the Lyapunov exponents of the isolated map ordered as $h_1 \ge h_2
\ge \dots \ge h_M$. The mode $l=0$  corresponds to the synchronized   state and
the other modes represents its transverse variations.  Hence $\mu_1(l)$ gives
the largest transverse Lyapunov exponent for the mode $l \neq 0$.  Therefore
the stability of synchronized state is ensured if $\mu_1(l) < 0$ for all $l
\neq 0$.  However, the symmetry in Fourier modes reduces this condition as
$\mu_1(l) < 0$ for $l=1,2,\dots,L/2$ ($(L-1)/2$ if $L$ is odd). Thus the
stability condition reduces to
\begin{align}
\label{stability_1d_pn_0}
\left|1-\frac{2}{P}\sum^P_{p=1} a_p \text{sin}^2(\pi pl/L)\right|<\text{exp}(-h_1), \\
l=1,2,\dots,L/2 \;\; \text{or} \;\; (L-1)/2, \nonumber
\end{align}
and this expression is also obtained in Ref.~\cite{rangarajan:03:01}.  We
use this condition to derive the expression for the critical lattice size
limit in the rest of this paper.

Let $\lambda_l =  1-\frac{2}{P}\sum^P_{p=1} a_p \text{sin}^2(\pi pl/L)$ and
define $\lambda_{\rm max} = \max \{\lambda_l\}$,   $\lambda_{\rm min} = \min
\{\lambda_l\}$. Then the above stability condition can be rewritten as:
\begin{align} \label{stability_1d_pn_02} \lambda_{\rm max} < \exp(-h_1), \ \
\lambda_{\rm min} > -\exp(-h_1). \end{align}

Let $\lambda^{\epsilon}_l =  1-\frac{2 \epsilon}{P}\sum^P_{p=1}
\text{sin}^2(\pi pl/L)$, $\epsilon = \min \{a_p\}$, and define an upper bound
on $\lambda_{\rm max}$ as $\lambda^*_{\rm max} = \max
\{\lambda^{\epsilon}_l\}$. Therefore the first stability condition is ensured
if $\lambda^*_{\rm max} < \exp(-h_1)$.  Similarly, let $\lambda^{\epsilon'}_l
=  1-\frac{2 \epsilon'}{P}\sum^P_{p=1} \text{sin}^2(\pi pl/L)$, $\epsilon' =
\max \{a_p\}$ and define a lower bound on $\lambda_{\rm min}$ as
$\lambda^*_{\rm min} = \min \{\lambda^{\epsilon'}_l\}$. Hence the second
stability condition in (\ref{stability_1d_pn_02}) is ensured if $\lambda^*_{\rm
min} > -\exp(-h_1)$.
\begin{figure}
\centering{\includegraphics[width=0.9\linewidth]{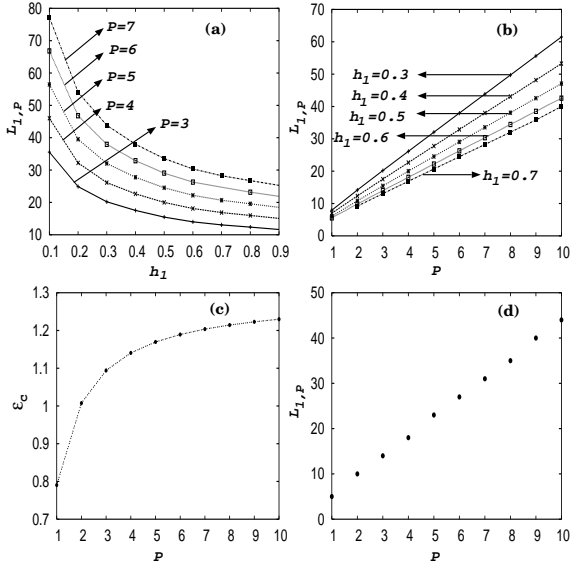}}
\caption{(a) The variation of $L_{1,P}$ with $h_1$,  (b) the variation of
$L_{1,P}$ with $P$, (c) the variation of $\epsilon_c$ with $P$  in one
dimensional coupled logistic map lattices, obtained for $r=1.9$, and (d) the
variation of $L_{1,P}$ with $P$ in one dimensional coupled logistic map lattices,
obtained for $r=1.9$ ($h_1=0.5554$).}
\label{fig_1d}
\end{figure}

We start with the simplest case of nearest neighbor coupling for which results
already exist \cite{rangarajan:02:01,chen:03:01} and then we extend
the idea to a more general $P$-neighbor coupling. In the nearest neighbor
coupling case, $\lambda_{\rm max}^{*} = 1-2\epsilon\, \text{sin}^2(\pi /L)$.
Therefore the first stability condition is satisfied if $\epsilon >
\frac{1-\text{exp}(-h_1)}{2 \text{sin}^2(\pi/L)}$, where $\epsilon \leq a_p \
\forall p$. Consequently, in terms of the coupling coefficients, the above
condition becomes
\begin{align}
\label{stability_1d_nn_2}
a_p > \frac{1-\text{exp}(-h_1)}{2 \text{sin}^2(\pi/L)}, \ \ \forall p.
\end{align}
Similarly, $\lambda_{\rm min}^{*} = 1-2\epsilon$. Hence the second stability
condition is satisfied if  $\epsilon' < \frac{1+\text{exp}(-h_1)}{2}$, where
$\epsilon' \geq a_p \ \forall p$. In terms of coupling coefficients, we get
\begin{align}
\label{stability_1d_nn_4}
a_p < \frac{1+\text{exp}(-h_1)}{2}, \ \ \forall p.
\end{align}
Combining the two inequalities
(\ref{stability_1d_nn_2})\&(\ref{stability_1d_nn_4}), we get the well-known \cite{rangarajan:02:01,chen:03:01}
final stability condition as:
\begin{align}
\frac{1-\text{exp}(-h_1)}{2 \text{sin}^2(\pi/L)} <  a_p <\frac{1+\text{exp}(-h_1)}{2}, \ \ \forall p.
\label{stability_1d_nn_5}
\end{align}
As $L$ becomes larger, the above stability range becomes smaller and at a particular
critical value of $L$, the range shrinks to zero. Beyond this critical value of
$L$, the stability condition (\ref{stability_1d_nn_5}) is violated and hence the synchronized
state can never be stable. At this critical value of $L$, one can replace the inequalities
by equality signs in Eq.(\ref{stability_1d_nn_5}) and get
\begin{align}
\ L_{1,1}=\text{Int}\left[\frac{\pi}{\text{sin}^{-1}(\sqrt{{\rm tanh}(h_1/2)})}\right],
\label{size_limit_1d_nn}
\end{align}
where $L_{1,1}$ is the maximum lattice size that can support synchronized chaos in
an one-dimensional nearest neighbor diffusively coupled map lattice.

Let us now turn to the more general case of $P$-neighbors diffusively coupled map
lattices for which no previous analytical results exist. However, similar results do exist
for coupled oscillators \cite{barahona:02:01,belykh:04:01} but they are in terms of numerically
computed stability ranges whereas we give analytical expressions.  After making use of some simple trigonometric relations, the
expressions for $\lambda^{\epsilon}_l$ and $\lambda^{\epsilon'}_l$ take the forms
\begin{align}
\lambda^{\epsilon}_l= 1- \epsilon \left[ 1- \frac{\text{sin}(P\pi l/L)\text{cos}\big( (P+1)\pi l/L \big)}{P\text{sin}(\pi l/L)} \right],
\label{stability_1d_pn_1}
\end{align}
and
\begin{align}
\lambda^{\epsilon'}_l= 1- \epsilon' \left[ 1- \frac{\text{sin}(P\pi l/L)\text{cos}\big( (P+1)\pi l/L \big)}{P\text{sin}(\pi l/L)} \right],
\label{stability_1d_pn_2}
\end{align}
respectively, where $\epsilon \leq a_p \leq \epsilon' \ \forall p$. The expression inside the
square bracket takes its lowest value when $l=1$  and it takes its highest
value for the mode $l=l_h=$Int$[L_{1,1}/2]$, for all values of $P$.   Following
the same procedure as in the nearest neighbor case, we finally get
\begin{widetext}
\begin{align}
\frac{1-\text{exp}(-h_1)}{\left[1- \frac{\text{sin}(P\pi /L)\text{cos}\big( (P+1)\pi /L \big)}{P\text{sin}(\pi /L)}\right]}
< a_p  <
\frac{1+\text{exp}(-h_1)}{\left[1- \frac{\text{sin}(P\pi l_h/L)\text{cos}\big( (P+1)\pi l_h/L \big)}{P\text{sin}(\pi l_h/L)}\right]}
\ \ \forall p.
\label{stability_1d_pn_5}
\end{align}
\end{widetext}
At this critical coupling strength $\epsilon_c$ ($a_p$=$\epsilon_c$ $\forall
p$) the extremes values of $a_p$ coincide and the above expression becomes
\begin{widetext}
\begin{align}
&\frac{\text{sin}(P\pi /L_{1,P})\text{cos}\big( (P+1)\pi /L_{1,P} \big)}{\text{sin}(\pi/L_{1,P})}=
P\left[1-\text{tanh}(h_1/2)
\left(1-\frac{\text{sin}(P\pi l_h/L_{1,P})\text{cos}\big((P+1)\pi l_h/L_{1,P}\big)}{P\text{sin}(\pi l_h/L_{1,P})}\right)\right],
\label{size_limit_1d_pn}
\end{align}
\end{widetext}
where $l_h=$ Int$[L_{1,1}/2]$ and $L_{1,1}$ is given in
Eq.~(\ref{size_limit_1d_nn}).  The critical lattice size limit $L_{1,P}$ is
obtained by solving the above transcendental equation numerically.  In a
special case of $P=L/2$ (or $(L-1)/2$ for odd $L$) we get
$h_1=2\,$tanh$^{-1}(1)$.  This result indicates that the synchronized state is
always possible for globally coupled map lattices as long as $h_1<\infty$.
The dependence of $L_{1,P}$ on the maximum Lyapunov exponent of the isolated
map ($h_1$) and  the number of neighbors coupled ($P$) are shown in
Fig.~\ref{fig_1d}(a) and \ref{fig_1d}(b). It is observed that $L_{1,P}$ increases
almost linearly with $P$ for a particular value of $h_1$, and decays with $h_1$
for a particular value of $P$. Also, all the results are confirmed numerically
by considering the logistic map $\big($defined by $x(n+1)=1-r[x(n)]^2$$\big)$
at each node. The variations of the critical coupling strength $\epsilon_c$ and
the critical size limit $L_{1,P}$ with $P$ (for $r=1.9$) are shown in
Fig.~\ref{fig_1d}(c) and \ref{fig_1d}(d).

\section{Critical Size Limit in 2D case}
\label{stability_2d}
Now we consider 2-dimensional coupled map lattices with $P$-neighbor diffusive
coupling of the form
\begin{widetext}
\begin{align}
{\bf x}_{j,k}(n+1) =   f\big({\bf x}_{j,k}(n)\big) + \frac{1}{4P} \sum^P_{p=1} \Big\{a_p \left[ f\big({\bf x}_{j-p,k}(n)\big) +
f\big({\bf x}_{j+p,k}(n)\big) - 2f\big({\bf x}_{j,k}(n)\big) \right]+   \nonumber  \\    b_p  \left[ f\big({\bf x}_{j,k-p}(n)\big) +
f\big({\bf x}_{j,k+p}(n)\big) - 2f\big({\bf x}_{j,k}(n)\big) \right]  \Big\},
\label{cml_2d}
\end{align}
\end{widetext}
where ${\bf x}_{j,k}$ is a $M$-dimensional state vector, $(j,k)$ represents the
lattice sites, $L$ is the lattice size, and $a_p$ and $b_p$ are the coupling
strengths between $(j,k)$th map and its $p$th neighbor along $j$ and $k$
directions, respectively.

In this case, the stability condition for synchronized state is
\begin{align}
\Big|1-\frac{1}{P}\sum^P_{p=1}\big[ a_p\text{sin}^2(\pi pl/L)
+b_p\text{sin}^2(\pi pm/L)\big]\Big| \nonumber  \\
< \exp(-h_1),
\label{trans_lya_2d_2}
\end{align}
where $l,m = 0,1, \ldots ,L-1,\ (l,m) \neq (0,0)$.

If we define $\nu_{l,m} = 1-\frac{1}{P}\sum^P_{p=1}\big[ a_p\text{sin}^2(\pi
pl/L)+b_p\text{sin}^2(\pi pm/L)\big]$,   $(l,m) \neq (0,0)$,   $\nu_{\rm max} =
\max \{\nu_{l,m}\}$,  and $\nu_{\rm min} = \min \{\nu_{l,m}\}$ then the above
stability condition becomes:
\begin{align}
\label{stability_2d_pn_02}
\nu_{\rm max} < \exp(-h_1), \ \
\nu_{\rm min} > -\exp(-h_1).
\end{align}

Performing an analysis similar to the 1D case, we obtain the expression for
critical lattice size limit for nearest neighbor coupling as
\begin{align}
\ L_{2,1}=\text{Int}\left[\frac{\pi}{\text{sin}^{-1}(\sqrt{2 {\rm tanh}(h_1/2)})}\right].
\label{size_limit_2d_nn}
\end{align}

\begin{figure}
\centering{\includegraphics[width=0.9\linewidth]{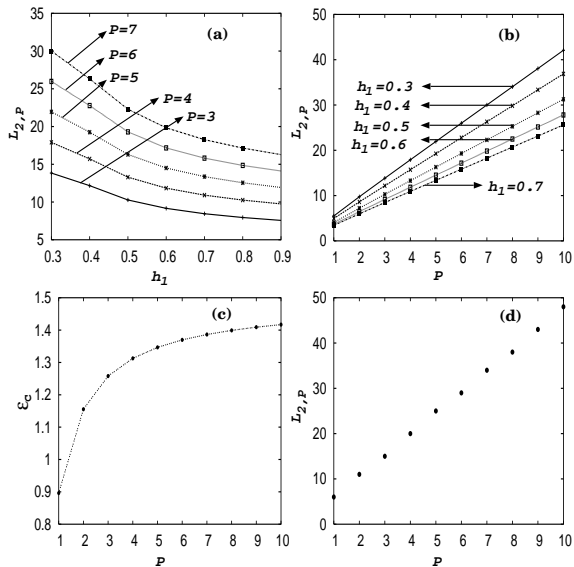}}
\caption{(a) The variation of $L_{2,P}$ with $h_1$,  (b) the variation of
$L_{2,P}$ with $P$, (c) the variation of $\epsilon_c$ with $P$  in
$2$-dimensional coupled logistic map lattices, obtained for $r=1.5$, and (d) the
variation of $L_{2,P}$ with $P$ in $2$-dimensional coupled logistic map lattices,
obtained for $r=1.5$ ($h_1=0.2378$).}
\label{fig_2d}
\end{figure}

In the case of $P$-neighbor diffusive coupling, we obtain the expression for
the critical lattice size limit $L_{2,P}$  as
\begin{widetext}
\begin{align}
&\frac{\text{sin}(P\pi /L_{2,P})\text{cos}\big( (P+1)\pi /L_{2,P} \big)}{\text{sin}(\pi/L_{2,P})}=
P\left[1-\text{2tanh}(h_1/2)
\left(1-\frac{\text{sin}(P\pi l_h/L_{2,P})\text{cos}\big((P+1)\pi l_h/L_{2,P}\big)}{P\text{sin}(\pi l_h/L_{2,P})}\right)\right],
\label{size_limit_2d_pn}
\end{align}
\end{widetext}
where $l_h=$Int$[L_{2,1}/2]$, \; $L_{2,1}$ is the critical lattice size limit
for $2$-dimensional nearest neighbor diffusively coupled map lattices and is
given in Eq.~(\ref{size_limit_2d_nn}).  $L_{2,P}$ is obtained by solving
Eq.~(\ref{size_limit_2d_pn}) numerically.  The dependence of $L_{2,P}$ on the
maximum Lyapunov exponent of the isolated map ($h_1$) and the number of
neighbors coupled ($P$) are shown in Fig.~\ref{fig_2d}(a) and \ref{fig_2d}(b).  In
our numerical verification, we have again considered logistic map at each
node.  The variations of the critical coupling strength $\epsilon_c$ and the
critical size limit $L_{2,P}$ (for $r=1.5$) with $P$  are shown in
Fig.~\ref{fig_2d}(c) and \ref{fig_2d}(d).

\section{Conclusions}
\label{con}
We have presented expressions for the critical lattice size limits ($L_{1,P}$
and $L_{2,P}$) for both $1$ and $2$-dimensional coupled map lattices with $P$
neighbor coupling. In both the cases, the value of these critical size limits
increase almost linearly with the number of coupled neighbors. In addition, all
the above results were verified through numerical studies using coupled
logistic map lattices.  Moreover, as $P$ increases to the global coupling limit
we showed explicitly that the critical size limit tends to infinity. However, our results are not valid for
discontinuous maps (for example, tent and Bernoulli maps considered in
Refs.~\cite{ginelli:03:01, cencini:05:01}) since the linear stability
analysis used in this study requires the maps to be continuous.

\acknowledgments

This work was supported by grants from DRDO, UGC (under SAP-Phase IV). PP was also
supported by DST
(under FAST-TRACK Young Scientist Scheme). GR is associated with the Jawaharlal
Nehru Centre for Advanced Scientific Research, Bangalore as a Honorary Faculty
Member.


\end{document}